\documentclass[preprint]{article}
\usepackage{amsmath,amsthm,amssymb,amscd,amsfonts}
\usepackage{amsmath,amsthm,mathrsfs}
\usepackage[colorlinks=true]{hyperref}
\usepackage{graphicx}
\usepackage{dcolumn}
\usepackage{epstopdf}

\title{\textbf{Renyi Holographic dark energy and its behaviour in $f(G)$ gravity}}
\author{$^{1}$ Md Khurshid Alam,$^{2}$S. Surendra Singh and $^{3}$L. Anjana Devi \\
\{\emph{Email}:$^{1}$alamkhurshid96@gmail.com, $^{2}$ssuren.mu@gmail.com,\\
  $^{3}$anjnlam@gmail.com\}}
\date{\small $^{1,2,3}$Department of Mathematics, National Institute of Technology Manipur-795004, (India).\\}

\begin{document}
\maketitle{}

\begin{abstract}
In this work, the Renyi holographic dark energy (RHDE)and its behaviour has been explored with the anisotropic and spatially homogeneous Bianchi type-I Universe in the framework of  $f(G)$ gravity. We use IR cutoff as the Hubble and Granda-Oliveros (GO) horizons. To find a consistent solutions of the field equations of the models, it is assumed that the deceleration parameter is defined in terms of function of Hubble parameter $H$. With reference to current cosmological data, the behaviors of the cosmological parameters relating to the dark energy model are evaluated and their physical significance is examined. It is observed that for both the models, the equation of state parameter approaches to $-1$ at late times. However, the RHDE model with the Hubble horizon exhibits stability from the squared sound speed, but the RHDE model with the GO horizon exhibits instability. In both the models, deceleration parameter and statefinder diagnostic confirm the accelerated expansion of the Universe and also correspond to the $\Lambda$CDM model at late times.\\


\end{abstract}

\section{Introduction}\label{sec1}
 General relativity (GR) is regarded as a key theory to comprehend several complexities of gravitational influences that offer a fundamental explanation of astrophysical events as well as the cosmos. The most significant truth that the universe suffers early inflation as well as late-time accelerated expansion has been revealed by a number of observational findings in recent years {\cite{Perlmutter99,Riess98,Huang06,Eisenstein05,Fedeli09,Koivisto06}}. The exotic substance of extremely high negative pressure known as dark energy (DE) which is the cause of the universe's expansion at an accelerated rate that accounts for 68 percent of the known universe total density. Its nature continues to be a mystery still. The cosmological constant $(\Lambda)$, which Einstein incorporated into the field equations in General Relativity, provides the straightforward argument for DE. This cosmological constant is thought to be extremely compatible with the observational data and has an equation of state (EoS) parameter of $\omega=-1$. Some dynamic models of DE, such as quintessence {\cite{Caroll98,Turner22}}, phantom {\cite{Caldwell02}}, k-essence \cite{Chiba2000}, tachyons {\cite{Padmanabhan02}}, chaplygin gas {\cite{Kamenshchik01}}, etc, have been proposed in response to the challenges associated with its theoretically expected order of magnitude with respect to that of the vacuum energy {\cite{Zlatev99}}. Another category of dynamic DE models allow us to accelerate the expansion without introducing any form of energy. These categories are known as modified gravity theories, which give an accelerated expansion through a modification in the action such as $f(T)$ gravity, $f(R,T)$ gravity, $f(R,G)$ gravity, $f(T,T)$ gravity and $f(R,T,R_{\mu\nu}T^{\mu\nu})$ gravity where $T$ is the trace of the energy-momentum tensor, $R_{\mu\nu}$ is the Ricci tensor and $G$ is the Gauss-Bonnet (GB) invariant {\cite{Myrzakulov11a,Linder10,Laurentis15,Harko11}}. Modified Gauss-Bonnet (GB) gravity, also known as $f(G)$ gravity, is one of the modified forms of GR that uses an arbitrary function of $G$, a quadratic invariant of the Gauss-Bonnet equation in the Einstein-Hilbert action {\cite{Nojiri06b}}. The motivation for $f(G)$ theory is mostly based on string theory via low energy effective scale \cite{Cognola06}. This approach effectively explains the accelerated expansion of the Universe which change from the decelerating to accelerating phase, satisfactory system tests, essential for Sadjadi's explanation of thermodynamics \cite{Sadjadi11} and characterization of all possible four types of future singularities by Bamba et al. \cite{Bamba10}. Thus one can construct feasible and consistent general theory of relativity models with local constraints by using $f(G)$. Myrzakulov et al. \cite{Myrzakulov11} investigated this theory to examine the DE  as well as the inflationary era. The reconstruction scenario of the most recent agegraphic dark energy (NADE) model and the $f(G)$ theory within the flat FRW space-time was taken into consideration by Jawad et al. \cite{Jawad13}. Shamir \cite{Samir16} reviewed the anisotropic space-time in $f(G)$ gravity. Sharif and Fatima \cite{Sharif14} studied energy conditions in $f(G)$ theory. Shaikh et al. \cite{Shaikh20} LRS Bianchi type-I models with HDE within $f(G)$ theory of gravity using different scale factors. Koussour et al. \cite{Koussour22} compared HDE in $f(G)$  gravity within Bianchi type-I space-time with the $\Lambda$CDM model by analysing the jerk parameter.  \\*
Particularly among the different dynamical DE models, the HDE mode has recently emerged as an effective method for researching the DE riddle. It was put forth based on the quantum characteristics of black holes (BH), which have been thoroughly studied in the literature to research quantum gravity {\cite{Li04,Susskind95}}. By holographic principle, we know that in a system with size $L$, bound on the vacuum energy $(\Lambda)$ must be under the limit of same size of the BH mass because of the formation of BH in quantum field theory. The energy density of HDE is defined as  $\rho_{\Lambda}=3d^{2}m_{p}^{2}L^{-2}$ where $m_{p}$ is the reduced Planck mass, $3d^{2}$ numerical constant and $L$ is IR-cutoff (Cohen et al. {\cite{Cohen99}}). In the literature, various types of IR-cutoff have been investigated, for example Hubble horizon $H^{-1}$, particle horizon, event horizon, Ricci scalar radius, conformal universe age  and Granda-Oliveros cutoff {\cite{Gao06,Granda08,Granda09,Wei08}}. Several HDE models with different IR-cutoffs may provide the recent accelerated expansion of the universe and demonstrate that transition from early decelerated epoch $(q>0)$ to current accelerated epoch $(q<0)$ is in consistent with recent observational data. It can also resolve the problem of cosmic coincidence {\cite{Pavon07}}. Number of studies suggested that the HDE model and observational data are in a fair amount of agreement {\cite{Xu10,Zhang09,Wang10,Duran11}}. By using generalized HDE and phantom cosmology, Nojiri and Odintsov {\cite{Nojiri06}} suggested a method to unify the early phase as well as late-time epochs of universe, and they also advocate for generalized concept as Hinflation {\cite{Nojiri19}}. Based on several formalism of entropy,HDE models  are formulated such as Tsallis HDE (THDE)\cite{Tsallis13}, Sharma-Mittal HDE (SMHDE) \cite{Jahromi18} and Renyi HDE model (RHDE) \cite{moradpour18}. THDE model is unstable at the classical level, whereas SMHDE and RHDE are stable in the case of non-interacting cosmos. Prasanthi and Aditya \cite{Prasanthi20} studied RHDE in Bianchi type $VI_{0}$ space-time and found that the Hubble cutoff is stable whereas the Granda-Oliveros cutoff is unstable. They also constrained the observational values of RHDE in Kantowski-Sachs universe \cite{Prasanthi21}. Shekh \cite{Shekh21} studied holographic and Renyi holographic dark energy models with the help of FLRW line element in $f(Q)$ gravity. \\*
Since anisotropy was crucial in the early stages of cosmic evolution, the anisotropic universe has recently caught the interest of many physicists. Additionally, the cosmic microwave background (CMB) anomalies from the Planck data \cite{Nojiri06b}, which were acquired, supported the notion of an anisotropy phase at the beginning of the Universe followed by an isotropy phase. The Bianchi type-I model has been examined by a number of researchers \cite{sarat19a,sarat19b,Anjana22,Khurshid22}. Based on the aforementioned studies, we investigate the Renyi holographic model of DE with $f(G)$ gravity in the Bianchi type-I universe in this paper. In order to solve the field equations and determine various physical variables, we shall assume that the deceleration parameter (DP) is a function of the Hubble parameter $H$. Following is the breakdown of the paper: The introduction is found in Sect. \ref{sec1}. We construct the action of $f(G)$ gravity and the field equation in Sect. \ref{sec2}. We have developed the Bianchi type-I metric and provided a few physical and geometrical parameters in Sect. \ref{sec3}. In Sect. \ref{sec4}, we studied the models of Renyi holographic dark energy . Sect. \ref{sec5} explain about the cosmological parameters. A conclusion is included in the final section \ref{sec6}.

\section{Formulation of Gauss-Bonnet gravity}\label{sec2}

The $f(G)$ gravity's modified Einstein-Hilbert action is configured {\cite{nojiri05}}as follows
\begin{equation}\label{1}
S=\frac{1}{2\kappa^{2}}\int d^{4}x\sqrt{-g}[R+f(G)]+S_{M}(g^{\mu \nu},\psi).
\end{equation}
In this case, $g$ denotes the determinant of metric tensor $g^{\mu \nu}$, $\kappa$ is the coupling constant, $f(G)$ is a general differentiable function of GB invariant, $R$ is the Ricci scalar, $S_{M}$ stands for a matter action which is a function of a space time metric $g_{\mu \nu}$ and matter fields $\psi$. The equation of invariant GB quantity is given as
\begin{equation}
G=R^{2}-4R_{\mu \nu}R^{\mu \nu}+4R_{\mu \nu \alpha \beta}R^{\mu \nu \alpha \beta}.
\end{equation}
By varying the action (\ref{1}) w. r. t. $g_{\mu \nu}$ shows the resulting equation
\begin{eqnarray}\label{2}\nonumber
G_{\mu \nu}+8\bigg[R_{\mu \rho \nu \sigma}+R_{\rho \nu}g_{\sigma \mu}-R_{\rho \sigma}g_{\nu \mu}-R_{\mu \nu}g_{\sigma \rho}+R_{\mu \sigma}g_{\nu \sigma}\\
+\frac{1}{2}R(g_{\mu \nu}g_{\sigma \rho}-g_{\mu \sigma}g_{\nu \rho})\bigg]\nabla^{\rho}\nabla^{\sigma}f_{G}+(Gf_{G}-f)g_{\mu \nu}=\kappa^{2}T_{\mu \nu},
\end{eqnarray}
where $\nabla_{\mu}$ denotes covariant differentiation, the Einstein tensor, $G_{\mu \nu}=R_{\mu \nu}-\frac{1}{2}Rg_{\mu \nu}$, $T_{\mu \nu}$ is the usual energy momentum tensor of matter fluid and $f_{G}$ stands for the derivation of $f$ with respect to $G$.
\\

\section{ Field equations and solutions}\label{sec3}

As observations highlight the possibility of anisotropic behavior of universe, the geometry of the spatially homogenous and anisotropic Bianchi type-I space-time, represented by the following metric is considered
\begin{equation}\label{3}
ds^{2}=dt^{2}-A^{2}(t)dx^{2}-B^{2}(t)(dy^{2}+dz^{2}).
\end{equation}
Here $A$ and $B$ are time dependent functions. Thus for this LRS Bianchi type-I metric, the Ricci scalar $R$ and GB invariant are respectively obtained as
\begin{equation}\label{4}
R=-2\bigg[\frac{\ddot{A}}{A}+2\frac{\ddot{B}}{B}+2\frac{\dot{A}}{A}\frac{\dot{B}}{B}+\frac{\dot{B}^{2}}{B^{2}} \bigg],
\end{equation}
\begin{equation}\label{5}
G=8\bigg[\frac{\ddot{A}\dot{B}^{2}}{AB^{2}}+2\frac{\dot{A}\dot{B}\ddot{B}}{AB^{2}} \bigg].
\end{equation}
The matter and holographic dark energy have the energy momentum tensors in the form
\begin{equation}\label{6}
T_{\mu \nu}=\rho_{m}u_{\mu}u_{\nu},
\end{equation}
and
\begin{equation}\label{7}
\widetilde{T_{\mu \nu}}=(\rho_{\Lambda}+p_{\Lambda}){\mu}u_{\nu}+g_{u_{\mu}u_{\nu}}p_{\Lambda}.
\end{equation}
where $\rho_{m}$ and $\rho_{\Lambda}$ are the energy densities of matter and holographic dark energy respectively and $p_{\Lambda}$ is the pressure of the HDE.
 In this Bianchi type -I metric(\ref{3}), the field equations (\ref{2}) with the (\ref{6}) and (\ref{7}) give us the system of field equations given below
\begin{equation}\label{8}
-2\frac{\ddot{B}}{B}-\frac{\dot{B}^{2}}{B^{2}}+16\frac{\dot{B}\ddot{B}}{B^{2}}\dot{f_{G}}+8\frac{\dot{B}^{2}}{B^{2}}\ddot{f_{G}}-Gf_{G}+f=\kappa^{2}p_{\Lambda},
\end{equation}
\begin{equation}\label{9}
-\frac{\ddot{A}}{A}-\frac{\ddot{B}}{B}-\frac{\dot{A}\dot{B}}{AB}+8\bigg(\frac{\dot{A}\dot{B}}{AB}+\frac{\ddot{A}\dot{B}}{AB}\bigg)\dot{f_{G}}
+8\frac{\dot{A}\dot{B}}{AB}\ddot{f_{G}}-Gf_{G}+f=\kappa^{2}p_{\Lambda},
\end{equation}
\begin{equation}\label{10}
2\frac{\dot{A}\dot{B}}{AB}+\frac{\dot{B}^{2}}{B^{2}}-24\frac{\dot{A}\dot{B}^{2}}{AB^{2}}\dot{f_{G}}+Gf_{G}-f=\kappa^{2}(\rho_{m}+\rho_{\Lambda}),
\end{equation}
As we know, a dot $(\cdot)$ denote the derivation of the time $(t)$. The average scale factor $a(t)$ and the spatial volume $V$ are defined by
\begin{equation}\label{16}
V=a^{3}=AB^{2}.
\end{equation}
The general form of average Hubble parameter $H$ is defined as
\begin{equation}
H=\frac{\dot{a}}{a}=\frac{1}{3}(H_{1}+2H_{2}).
\end{equation}
Here $H_{1}=\frac{\dot{A}}{A}$ and $H_{2}=H_{3}=\frac{\dot{B}}{B}$ are directional Hubble parameter along $x$, $y$ and $z$ axes respectively.\\*
The continuity equation can be obtained as
\begin{equation}\label{18}
\dot{\rho_{m}}+\dot{\rho_{\Lambda}}+3H(\rho_{m}+\rho_{\Lambda}+p_{\Lambda})=0.
\end{equation}
The continuity equations of the matter and HDE are respectively obtained as
\begin{equation}\label{19}
\dot{\rho_{m}}+3H\rho_{m}=0.
\end{equation}
And
\begin{equation}\label{20}
\dot{\rho_{\Lambda}}+3H(\rho_{\Lambda}+p_{\Lambda})=0.
\end{equation}
 Applying the relation $p_{\Lambda}=\omega_{\Lambda}\rho_{\Lambda}$, the barotropic equation of state, the EoS HDE parameter can be found from (\ref{20}) as
\begin{equation}\label{21}
\omega_{\Lambda}=-1-\frac{\dot{\rho_{\Lambda}}}{3H\rho_{\Lambda}}.
\end{equation}
In this work, we assume that the function $f(G)$ obeys the power law models introduced by Cognola et al. {\cite{Cognola06}}
\begin{equation}\label{28}
f(G)=\eta G^{n+1},
\end{equation}
where $\eta$ and $n$ are arbitrary constants. The possibility of disappearing Big-Rip singularity and the ability to anticipate the occurrence of a transient phantom epoch that is consistent with the observational data are the main factors for choosing this power law  $f(G)$ model. For the Bianchi type-I universe (\ref{3}), deceleration parameter $(q)$, the scalar expansion $(\theta)$, the shear scalar $(\sigma^{2})$ and the average anisotropy parameter $(A_{m})$ have the form
\begin{equation}
q=-\frac{a\ddot{a}}{\dot{a}^{2}}=\frac{d}{dt}\bigg(\frac{1}{H}\bigg)-1,
\end{equation}
\begin{equation}
\theta=3H=\frac{\dot{A}}{A}+2\frac{\dot{B}}{B},
\end{equation}
\begin{equation}
\sigma^{2}=\frac{1}{2}\bigg[\sum_{i=1}^{3} H_{i}^{2}-3H^{2}\bigg],
\end{equation}
\begin{equation}
A_{m}=\frac{1}{3}\sum_{i=1}^{3}\bigg(\frac{H_{i}-H}{H}\bigg)^{2}.
\end{equation}
Here, we take into account the expansion scalar $(\theta)$ is directly proportional to the shear scalar $(\sigma)$, which results for following relationship between the metric potentials:
\begin{equation}\label{11}
A=B^{m},
\end{equation}
Here, positive constant ($m$) accounts for the anisotropic evolution of space time. When $m=1$, the model is isotropic; else it is anisotropic. Logic behind this condition is described with reference to {\cite{thorne67}}. Observational evidence indicates the current isotropic expansion of universe by about $\approx 30\%$ \cite{Kristian66}. More specifically, redshift studies set the limit at $\frac{\sigma}{H}\leq0.3$, in the neighbourhood of our present day galaxy. According to Collins et al.\cite{Collins77},  the normal congruence follows the above condition ($\frac{\sigma}{H}$ is constant) for a spatially homogenous space-time. In accordance with recent data, we are also interested in finding an acceptable cosmological explanations that show a transition from early deceleration to late acceleration. To solve this problem, a number of different assumptions can be used. Observations demonstrate the advance of Universe through a phase change from the earlier decelerating expansion to the present accelerating one, which is the reason for accounting for the time-dependent deceleration parameter  ($q$). The $q$ is a geometrical parameter that, depending on its sign, depicts the universe acceleration or deceleration. For this scenario, we understand that the universe experiences accelerating expansion for $q<0$; when $q>0$, the universe experiences decelerating expansion; when $q=0$, constant expansion of universe is shown whereas $q<-1$ stands for super-exponential expansion. As a result of the foregoing, we decided to use  deceleration parameter $q$ as a function of the Hubble parameter $H$ as proposed by Tiwari {\cite{tiwari20}}
\begin{equation}\label{12}
q=\alpha-\frac{\beta}{H},
\end{equation}
where $\alpha$ and $\beta$ are constants. The desired transition from positive to negative is achieved by this form of the deceleration parameter. The scale factor and Hubble parameter can be calculated using equation (\ref{12}) as follows
\begin{equation}\label{13}
a=k_{1}(e^{\beta t}-1)^{\frac{1}{1+\alpha}},
\end{equation}
where $k_{1}$ is the integration constant. From equation (\ref{13}), in order to have an expanding Universe, we can deduce that $\alpha>-1$, $\beta>0$. Also the scale factors vanishes at $t=0$, hence our model has a point type singularity at the early Universe.
From this above equation, we can immediately derive the spatial volume as $V=k_{1}^{3}(e^{\beta t}-1)^{\frac{3}{1+\alpha}}$, which has value zero in the beginning and increases with increase of $t$, which shows that our model is expanding with time. And
\begin{equation}\label{14}
H=\frac{\beta e^{\beta t}}{(1+\alpha)(e^{\beta t}-1)}.
\end{equation}
From this equation, we can understand that at the beginning, $H$ is infinite and with the passage of time it decreases to a constant value $\frac{\beta}{1+\alpha}$.
Using equations (\ref{11}) and (\ref{13}) in equation (\ref{16}), the metric potentials $A$ and $B$ are found as
\begin{equation}
A=k_{1}^{3m}(e^{\beta t}-1)^{\frac{3m}{(1+\alpha)(m+2)}},
\end{equation}
\begin{equation}
B=k_{1}^{3}(e^{\beta t}-1)^{\frac{3}{(1+\alpha)(m+2)}}.
\end{equation}

\begin{figure}
\begin{center}
\includegraphics[scale=0.8]{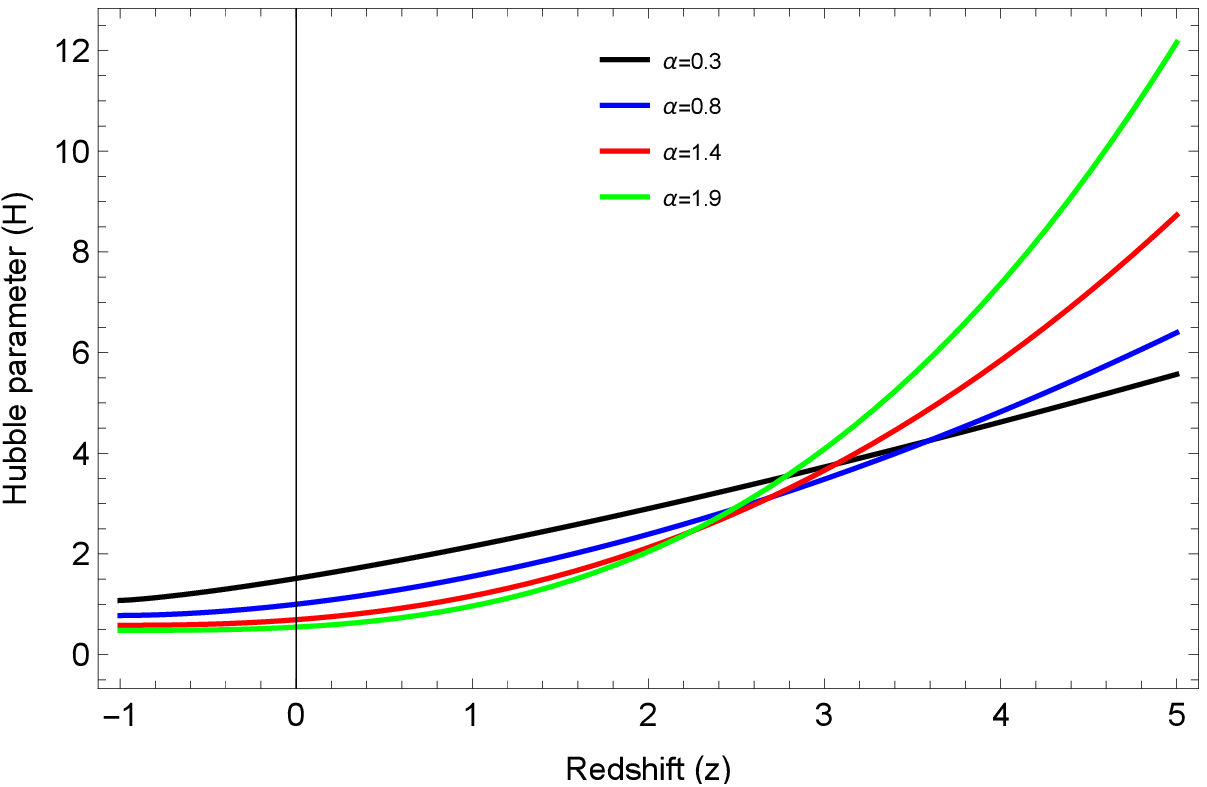}

\hspace{1cm}\vspace{5mm}
\footnotesize{Fig. 1. Hubble parameter ($H$) versus redshift ($z$) for $k_{1}=0.5$, $\beta=1.4$ and $\alpha=0.3,0,8,1.4,1.9$.}
\vspace{3mm}
\end{center}
\end{figure}

\begin{figure}
\begin{center}
\includegraphics[scale=0.8]{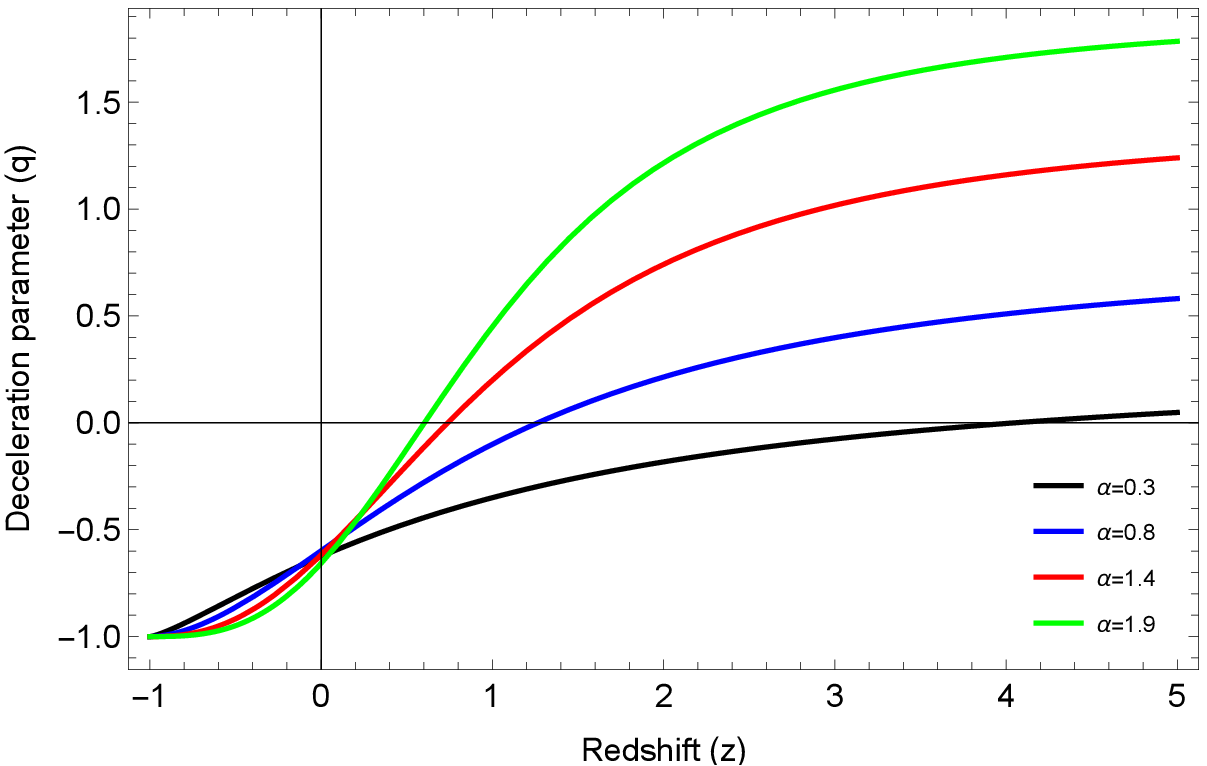}

\hspace{1cm}\vspace{5mm}
\footnotesize{Fig. 2. Deceleration parameter ($q$) versus redshift ($z$) for $k_{1}=0.5$, and $\alpha=0.3,0,8,1.4,1.9$.}
\vspace{3mm}
\end{center}
\end{figure}

With the use of above metric potentials, the metric (\ref{3}) can now be expressed as
\begin{equation}\label{17}
ds^{2}=dt^{2}-\bigg[k_{1}(e^{\beta t}-1)^{\frac{1}{(1+\alpha)(m+2)}}\bigg]^{6m}dx^2-\bigg[k_{1}(e^{\beta t}-1)^{\frac{1}{(1+\alpha)(m+2)}}\bigg]^{6}(dy^{2}+dz^{2}).
\end{equation}
Equation (\ref{17}) represents the spatially homogeneous and anisotropic Bianchi type-I RHDE model in the context of $f(G)$ gravity with the following properties together with the physical parameters described below. Using equation (\ref{14}) in equation (\ref{12}), we have
\begin{equation}\label{15}
q=-1+\frac{1+\alpha}{e^{\beta t}}.
\end{equation}
From this equation we can deduce that at the beginning, $q=\alpha$, a constant and with the increase of time, it approaches to $-1$ at late times, which shows that our model has a transition to acceleration.
The relation $a(t)=\frac{1}{1+z}$, where $z$ is the redshift, yield us the relationship as below
\begin{equation}
t(z)=\frac{1}{\beta}log\bigg[1+\frac{1}{\{k_{1}(1+z)\}^{1+\alpha}}\bigg].
\end{equation}
Additionally, redshift $(z)$ can be used to express the Hubble parameter $(H)$ as
\begin{equation}
H(z)=\frac{\beta}{1+\alpha}[1+\{k_{1}(1+z)\}^{1+\alpha}].
\end{equation}
Fig. 1 depicts the behavior of the Hubble parameter as a function of redshift at various $\alpha$ values  (i.e. $\alpha\geq0.3$). According to this graph, the Hubble parameter has a positive relationship with redshift. At the present, when $(z=0)$, the Hubble parameter is strictly positive, and for the early Universe, when $(z>0)$, it increases as $z$ increases. Also for $\alpha=1.4$, the current value of $H$ has been noted as $70.71~Kms^{-1}Mpc^{-1}$ which is in agreement with the observational value \cite{Planck18}.
Similarly, we get the deceleration parameter $(q)$ in terms of redshift $(z)$ as
\begin{equation}
q(z)=-1+\frac{(1+\alpha)\{k_{1}(1+z)\}^{1+\alpha}}{1+\{k_{1}(1+z)\}^{1+\alpha}}.
\end{equation}
The $q(z)$ exhibits two epochs throughout the universe: the initial deceleration phase and the current acceleration phase, as shown in Fig. 2, which depicts the parameter's behaviour in terms of redshift. In this study, $\alpha\geq0.3$ is required to produce both phases. The change from the initial deceleration phase to the present accelerated phase is also accomplished with a specific redshift, called the transition redshift $z$. According to the graph, the transition redshift for $\alpha=1.4$ is $z_{tr}=0.73$. Also the value of $q$ is found to be $-0.6$ in present time. Therefore, the results of our findings are in agreement with the observational values \cite{Planck18}.

The expressions of scalar expansion $(\theta)$, shear scalar $(\sigma^{2})$ and the average anisotropy parameter $(A_{m})$ are therefore obtained as
\begin{equation}\label{30}
\theta=\frac{3\beta e^{\beta t}}{(1+\alpha)(e^{\beta t}-1)},
\end{equation}
\begin{equation}\label{31}
\sigma^{2}=\frac{3\beta^{2} e^{2\beta t}}{(1+\alpha)^{2}(e^{\beta t}-1)^{2}}\frac{(m-1)^{2}}{(m+2)^{2}},
\end{equation}
\begin{equation}\label{32}
A_{m}=\frac{2(m-1)^{2}}{(m+2)^{2}}.
\end{equation}
From equations (\ref{30}) and (\ref{31}), we can deduce that  the scalar expansion and the shear scalar diverge at $t\rightarrow0$, then tends to respective constant values $\theta=\frac{3\beta }{1+\alpha}$ and $\sigma^{2}=\frac{3\beta^{2}(m-1)^{2}}{(1+\alpha)^{2}(m+2)^{2}}$ when $t\rightarrow \infty$. From equation (\ref{32}) it is observed that the anisotropic parameter remains constant during cosmic evolution which suggests that our model is uniformly anisotropic for $m\neq1$. We also observe from the equations (\ref{31}) and (\ref{32}) that when $m=1$, shear scalar $\sigma^{2}=0$ and anisotropic parameter $A_{m}=0$, the model becomes shear free and isotropic.

\begin{figure}
\begin{center}
\includegraphics[scale=1.1]{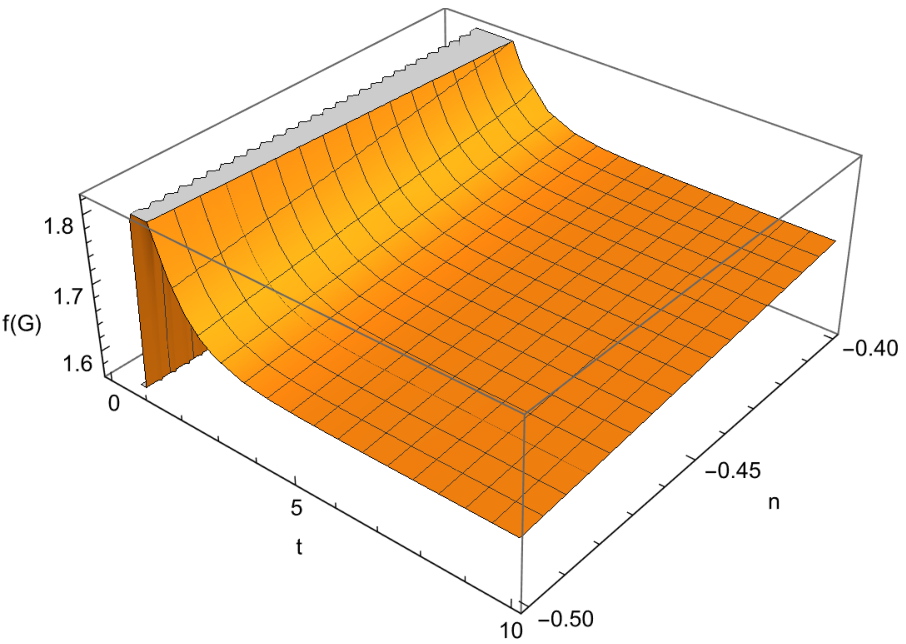}

\hspace{1cm}\vspace{5mm}
\footnotesize{Fig. 3. Evolution of $f(G)$ versus $n$ and $t$ for $\alpha=\beta=\eta=1.4$.}
\vspace{3mm}
\end{center}
\end{figure}

Also the GB invariant $G$ and Ricci scalar $R$ behave as
\begin{equation}\label{29}
G=\frac{648m\beta^{4}e^{3\beta t}}{(m+2)^{3}(1+\alpha)^{4}(e^{\beta t}-1)^{4}}\{e^{\beta t}-(1+\alpha)\},
\end{equation}
\begin{equation}
R=\frac{6\beta^{2}e^{2\beta t}}{(1+\alpha)^{2}(e^{\beta t}-1)^{2}}\bigg[\frac{1+\alpha}{e^{\beta t}}+\frac{3}{(m+2)^{2}}-3\bigg].
\end{equation}
Equations (\ref{28}) and (\ref{29}) are used to derive the function $f(G)$ given by
\begin{equation}
f(G)=\eta\bigg[\frac{648m\beta^{4}e^{3\beta t}}{(m+2)^{3}(1+\alpha)^{4}(e^{\beta t}-1)^{4}}\{e^{\beta t}-(1+\alpha)\} \bigg]^{n+1}.
\end{equation}
Fig. 3 depicts the $f(G)$ as a function of time $n<0$. It demonstrates that the function $f(G)$ has a transitory behaviour and is positive throughout cosmic time. $f(G)$ is very large at the beginning of evolution, approaches zero, then increases and ultimately takes a constant value as $\lim F(G) \rightarrow \eta\bigg[\frac{648m\beta^{4}}{(m+2)^{3}(1+\alpha)^{4}} \bigg]^{n+1}$ when $t\rightarrow\infty$.
\\
\section{R\'{e}nyi Holographic Dark Energy Models}\label{sec4}

We have consider a system with $n$ discrete states having probability distribution $P_{i}$ which satisfies the condition $\sum_{i=1}^{n} P_{i}= 1$. R\'{e}nyi entropy is a recognized generalized entropy parameter defined as \cite{jawad18}
\begin{equation}\label{22}
\mathcal{S}=\frac{1}{\delta}ln\sum_{i}^{n}P_{i}^{1-\delta}~~~,~~~S_{T}=\frac{1}{\delta}\sum_{i=1}^{n}(P_{i}^{1-\delta}-P_{i}),
\end{equation}
where $\delta\equiv1-U$ and $U$ is a real parameter and $T=\frac{1}{2\pi L}$  and $L$ is the IR cutoff. By using equation (\ref{22}), we obtain the relation
\begin{equation}\label{23}
\mathcal{S}=\frac{1}{\delta} ln(1+\delta S_{T}).
\end{equation}
In equation (\ref{23}), the Bekenstein entropy is given in the form $S_{T}=\frac{A}{4}$, where $A=4\pi L^{2}$. This gives the Renyi entropy of the system as
\begin{equation}
\mathcal{S}=\frac{1}{\delta} ln(1+\pi \delta L^{2}).
\end{equation}
Using the following assumption $\rho_{\Lambda}dV \propto TdS$, we can get RHDE as
\begin{equation}\label{24}
\rho_{\Lambda}=\frac{3d^{2}}{8\pi L^{2}}(1+\pi \delta L^{2})^{-1}.
\end{equation}

\begin{figure}
\begin{center}
\includegraphics[scale=0.8]{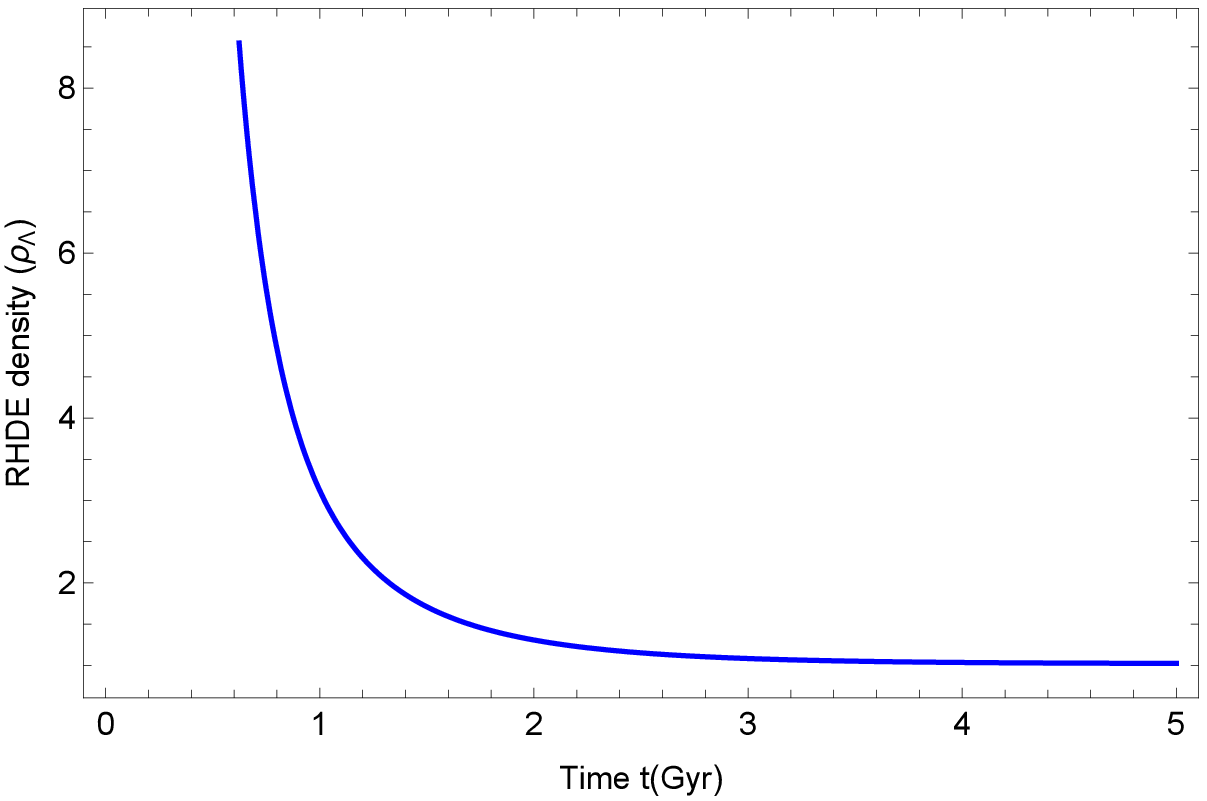}

\hspace{1cm}\vspace{5mm}
\footnotesize{Fig. 4. Holographic dark energy density ($\rho_{\Lambda}$) versus time ($t$)(Hubble horizon cutoff) for $\alpha=\beta=1.4$, $d=7$ and $\delta=5.2$.}
\vspace{3mm}
\end{center}
\end{figure}

\begin{figure}
\begin{center}
\includegraphics[scale=0.8]{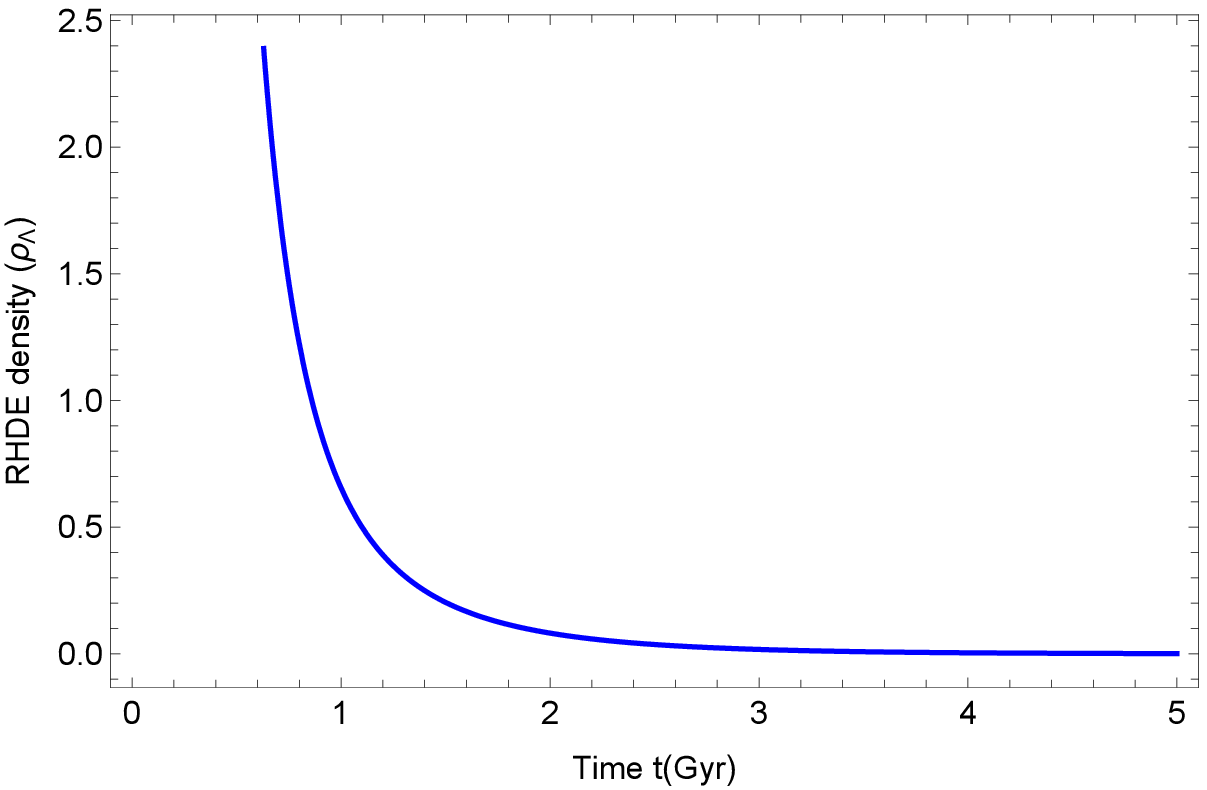}

\hspace{1cm}\vspace{5mm}
\footnotesize{Fig. 5. Holographic dark energy density ($\rho_{\Lambda}$) versus time ($t$)(GO cutoff) for $\alpha=\beta=1.4$, $d=7$, $\delta=5.2$, $\gamma_{1}=1.065$ and $\gamma_{2}=0.4$.}
\vspace{3mm}
\end{center}
\end{figure}
\subsection{Model-1: RHDE model with Hubble horizon cutoff}
Here, the Renyi holographic dark energy density is calculated by using the Hubble horizon as a candidate for the IR cutoff i.e. $L=H^{-1}$ and $8\pi=1$ is found to be
\begin{equation}\label{25}
\rho_{\Lambda}=\frac{3d^{2}H^{4}}{H^{2}+\pi \delta}.
\end{equation}
Using equation (\ref{14}) in equation (\ref{25}), we obtain energy density of RHDE in this model as
\begin{equation}\label{26}
\rho_{\Lambda}=\frac{3d^{2}(\beta e^{\beta t})^{4}}{\{\beta(1+\alpha)e^{\beta t}(e^{\beta t}-1)\}^{2}+\pi \delta\{(1+\alpha)(e^{\beta t}-1)\}^{4} }.
\end{equation}
From this expression for $\rho_{\Lambda}$, we can deduce that it is a positive decreasing function of time and when $t\rightarrow\infty$, it tends to a constant value $\frac{3d^{2}}{\beta^{2}(1+\alpha)^{2}+\pi\delta(1+\alpha)^{4}}$ which shows that this dark energy component will remain uniformly at late epoch. This phenomenon highlights the behavior of accelerated expansion of universe. Also using equation (\ref{14}) in (\ref{19}), we found the matter energy density as
\begin{equation}\label{33}
\rho_{m}=c_{1}k_{1}^{-3}(e^{\beta t}-1)^{\frac{-3}{1+\alpha}}.
\end{equation}

The coincidence parameter ($\bar{r}$) is defined as the ratio between the HDE density ($\rho_{\Lambda}$) and the matter energy density ($\rho_{m}$), therefore from equations (\ref{26}) and (\ref{33}) the coincidence parameter is found to be
\begin{equation}
\bar{r}=\frac{\rho_{\Lambda}}{\rho_{m}}=\frac{3d^{2}k_{1}^{3}}{ c_{1}}\frac{(\beta e^{\beta t})^{4}(e^{\beta t}-1)^{\frac{3}{1+\alpha}}}{\{\beta(1+\alpha)e^{\beta t}(e^{\beta t}-1)\}^{2}+\pi \delta\{(1+\alpha)(e^{\beta t}-1)\}^{4}}.
\end{equation}

The Renyi holographic dark energy density is plotted against time in Hubble's cut-off with appropriate values of constants as shown in Fig.4. It is shown that it remains positive and decrease with increase of time and the contribution of $\alpha$, $\beta$ and $\delta$ remains negligible in its behavior. From equation (\ref{33}), we can observed that the evolution of the matter energy density $(\rho_{m})$ begins with a positive value, but disappears later, which denotes the expansion of the Universe. It is also noted that the coincidence parameter $\bar{r}$ initially changes at a very early stage of development, but after a finite time, it converges to a constant value and stays constant throughout the evolution, avoiding the coincidence problem (unlike $\Lambda$CDM).
Equation of state parameter for RHDE in Hubble cutoff is
\begin{equation}\label{35}
\omega_{\Lambda}=-1+\frac{2}{3}\frac{(1+\alpha)[\beta^{2}e^{2\beta t}+2\pi\delta (1+\alpha)^{2}(e^{\beta t}-1)^{2}]}{e^{\beta t}[\beta^{2}e^{2\beta t}+\pi\delta (1+\alpha)^{2}(e^{\beta t}-1)^{2}]}.
\end{equation}
From this expression, we can deduce that the value of $\omega_{\Lambda}$ converges to $-1$ at late times, indicating the $\Lambda$CDM model, which coincides with observational data.
The RHDE pressure is obtained as
\begin{eqnarray}\nonumber
p_{\Lambda} &=& \frac{3d^{2}(\beta e^{\beta t})^{4}}{\{\beta(1+\alpha)e^{\beta t}(e^{\beta t}-1)\}^{2}+\pi \delta\{(1+\alpha)(e^{\beta t}-1)\}^{4} }\\
&& \times \bigg[-1+\frac{2}{3}\frac{(1+\alpha)[\beta^{2}e^{2\beta t}+2\pi\delta (1+\alpha)^{2}(e^{\beta t}-1)^{2}]}{e^{\beta t}[\beta^{2}e^{2\beta t}+\pi\delta (1+\alpha)^{2}(e^{\beta t}-1)^{2}]}\bigg].
\end{eqnarray}

\begin{figure}
\begin{center}
\includegraphics[scale=0.8]{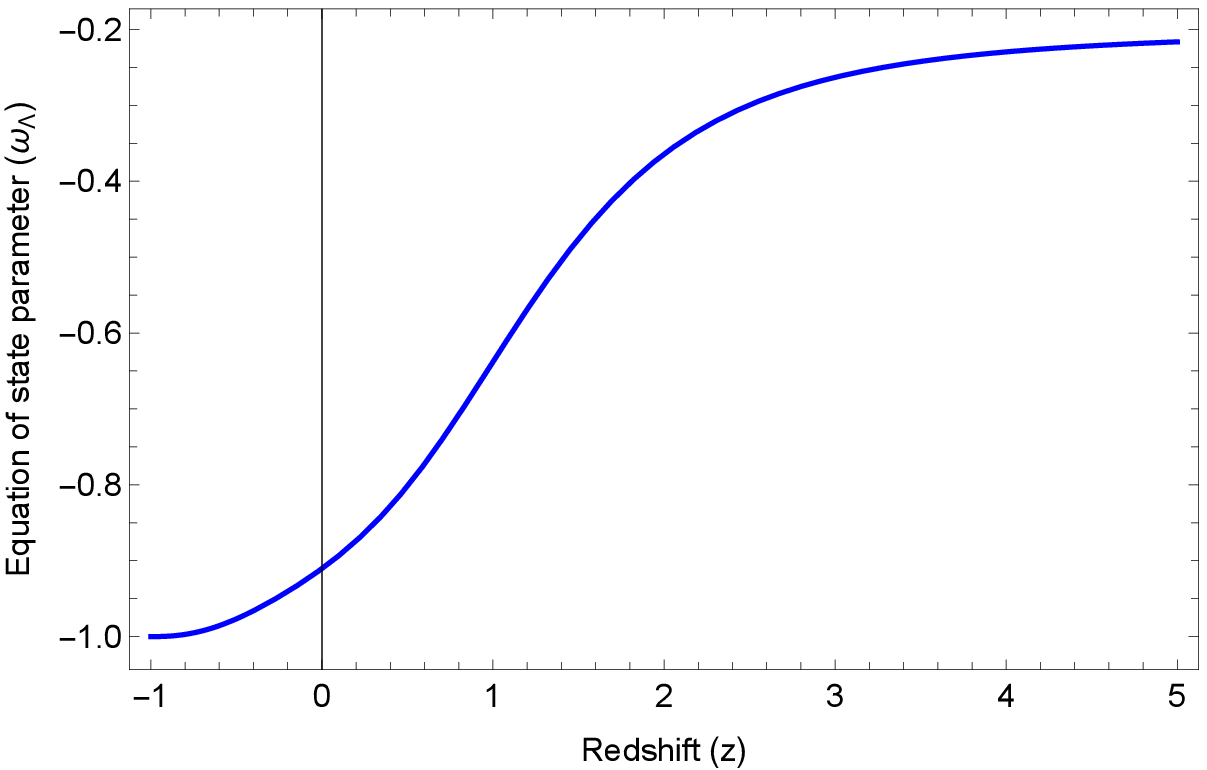}

\hspace{1cm}\vspace{5mm}
\footnotesize{Fig. 6. Equation of state ($\omega$) versus redshift ($z$) (Hubble horizon cutoff) for $k_{1}=0.5, \alpha=\beta=1.4$, $d=7$ and $\delta=5.2$.}
\vspace{3mm}
\end{center}
\end{figure}

\begin{figure}
\begin{center}
\includegraphics[scale=0.8]{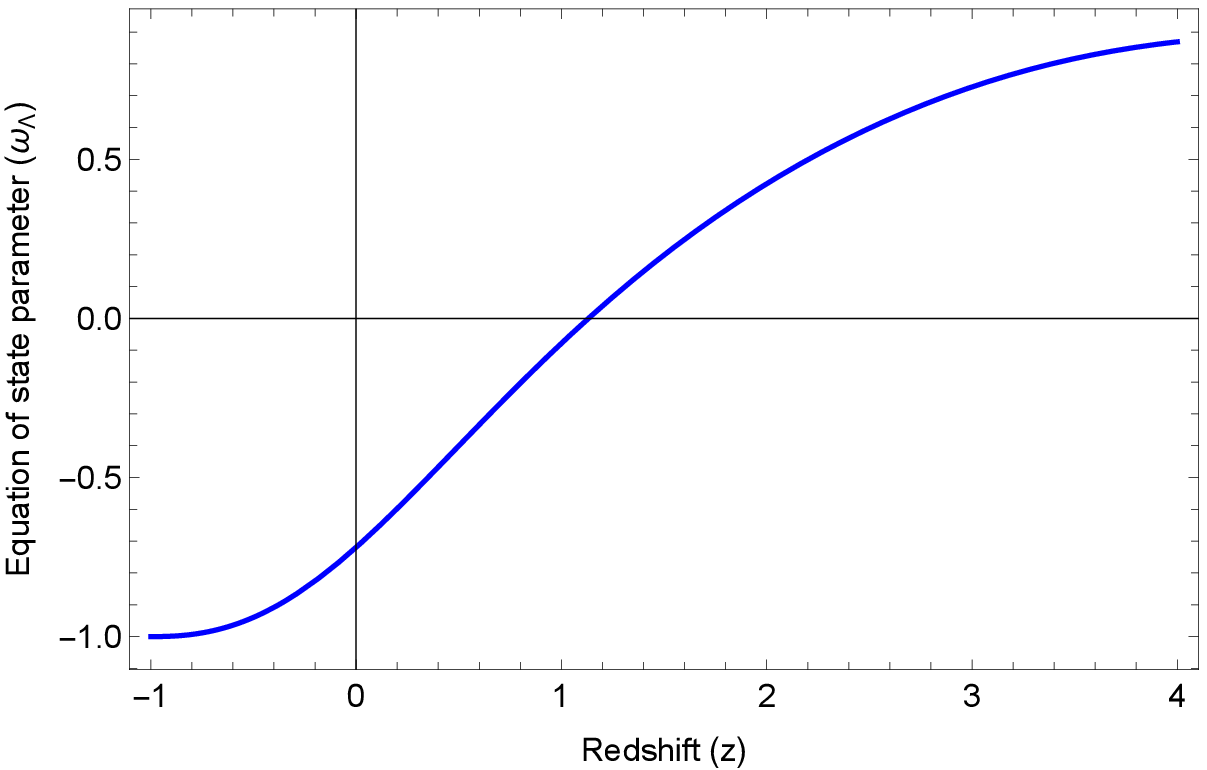}

\hspace{1cm}\vspace{5mm}
\footnotesize{Fig. 7. Equation of state ($\omega$) versus redshift ($z$) (GO cutoff) for $k_{1}=0.5, \alpha=\beta=1.4$, $d=7$, $\delta=5.2$, $\gamma_{1}=1.065$ and $\gamma_{2}=0.4$.}
\vspace{3mm}
\end{center}
\end{figure}
\subsection{Model-2: RHDE model with Granda-Oliveros horizon cutoff}
For this model, we consider RHDE model with GO horizon cut off i.e. $L=(\gamma_{1}H^{2}+\gamma_{2}\dot{H})^{-1/2}$ and $8\pi=1$. Substituting this value of $L$ in (\ref{24}), we have
\begin{equation}\label{27}
\rho_{\Lambda}=\frac{3d^{2}(\gamma_{1}H^{2}+\gamma_{2}\dot{H})^{2}}{\pi\delta+(\gamma_{1}H^{2}+\gamma_{2}\dot{H})}.
\end{equation}
Using equation (\ref{14}) in equation (\ref{27}), we found energy density of RHDE as
\begin{equation}\label{34}
\rho_{\Lambda}=\frac{3d^{2}\beta^{4} e^{2\beta t}\{\gamma_{1}e^{\beta t}-\gamma_{2}(1+\alpha)\}^{2}}{\pi\delta\{(1+\alpha)(e^{\beta t}-1)\}^{4}+e^{\beta t}\{\beta(1+\alpha)(e^{\beta t}-1)\}^{2}\{\gamma_{1}e^{\beta t}-\gamma_{2}(1+\alpha)\}}.
\end{equation}
From this expression, we can derive that the value of $\rho_{\Lambda}$ is very large in the beginning and decreases with the increase of time.
Also for this model, the matter energy density will be same as that of the RHDE with Hubble cutoff. Now from equations (\ref{26}) and (\ref{34}) the coincidence parameter becomes
\begin{equation}
\bar{r}=\frac{\rho_{\Lambda}}{\rho_{m}}=\frac{3d^{2}k_{1}^{3}}{ c_{1}}\frac{\beta^{4} e^{2\beta t}\{\gamma_{1}e^{\beta t}-\gamma_{2}(1+\alpha)\}^{2}(e^{\beta t}-1)^{\frac{3}{1+\alpha}}}{\pi\delta\{(1+\alpha)(e^{\beta t}-1)\}^{4}+e^{\beta t}\{\beta(1+\alpha)(e^{\beta t}-1)\}^{2}\{\gamma_{1}e^{\beta t}-\gamma_{2}(1+\alpha)\}}.
\end{equation}
The behavior of Renyi holographic dark energy density is plotted against time in Granda-Oliveros cutoff with the acceptable values of constant as shown in Fig 5. From the figure,  it is observed that the energy density of the model is constantly a positive function of time and decreases with increase of time. As we know from equation (\ref{33}), evolution of the matter energy density $(\rho_{m})$ starts at a positive value, but disappears at late times.  As in the RHDE Hubble cutoff, it is observed that the coincidence parameter $\bar{r}$ initially changes at a very early stage of development, but after a finite time, it converges to a constant value and stays constant throughout the evolution, avoiding the coincidence problem (unlike $\Lambda$CDM).\\*
Equation of state parameter for RHDE in Granda-Oliveros cutoff is
\begin{eqnarray}\label{36}\nonumber
\omega_{\Lambda} &=& -1+\frac{(1+\alpha)\{2\gamma_{1} e^{\beta t}-\gamma_{2}(1+\alpha)(e^{\beta t}+1)\}}{3e^{\beta t}\{\gamma_{1}e^{\beta t}-\gamma_{2}(1+\alpha)\}}\\
&&\times \frac{[2\pi\delta(1+\alpha)^{2}(e^{\beta t}-1)^{2}+\beta^{2}e^{\beta t}\{\gamma_{1}e^{\beta t}-\gamma_{2}(1+\alpha)\}]}{[\pi\delta\{(1+\alpha)(e^{\beta t}-1)\}^{2}+\beta^{2}e^{\beta t}\{\gamma_{1}e^{\beta t}-\gamma_{2}(1+\alpha)\}]}.
\end{eqnarray}
From this expression also, we can deduce that the value of $\omega_{\Lambda}$ converge to $-1$ at late times, indicating the $\Lambda$CDM model, which coincides with the observational data.
The RHDE pressure is obtained as
\begin{eqnarray}\nonumber
p_{\Lambda} &=& \rho_{\Lambda}\bigg[-1+\frac{(1+\alpha)\{2\gamma_{1} e^{\beta t}-\gamma_{2}(1+\alpha)(e^{\beta t}+1)\}}{3e^{\beta t}\{\gamma_{1}e^{\beta t}-\gamma_{2}(1+\alpha)\}}\\
&&\times \frac{[2\pi\delta(1+\alpha)^{2}(e^{\beta t}-1)^{2}+\beta^{2}e^{\beta t}\{\gamma_{1}e^{\beta t}-\gamma_{2}(1+\alpha)\}]}{[\pi\delta\{(1+\alpha)(e^{\beta t}-1)\}^{2}+\beta^{2}e^{\beta t}\{\gamma_{1}e^{\beta t}-\gamma_{2}(1+\alpha)\}]}\bigg].
\end{eqnarray}

\section{Cosmological Parameters}\label{sec5}
This section investigates how the universe expands using the cosmological parameters including equation of state (EoS),  squared sound speed $(v_{s}^{2})$,  density parameter ($\Omega$),  state finder parameter $(r,s)$ and the energy conditions for both the derived anisotropic RHDE models.
\subsection{EoS parameter}
The various phases of the expanding Universe are commonly categorised using the equation of state parameter $(\omega)$. Particularly, the transition between the decelerated and accelerated phases has phases where radiation and DE predominate. EoS parameter is defined as  $\omega=\frac{p}{\rho}$ where $p$ is pressure and $\rho$ is energy density of matter distribution.  The eras that make up the decelerated and accelerated phases are as follows: Decelerated phase (Cold dark matter or dust fluid $\omega=0$, radiation era $0<\omega<\frac{1}{3}$ and stiff fluid $\omega=1$) and Accelerated phase (Cosmological constant or vacuum era $\omega=-1$, quintessence $-1<\omega<\frac{-1}{3}$ and quintom era). Fig. 6 displays the graphical behavior of the Renyi holographic dark energy equation of state parameter versus time $t$ in Hubble's cutoff for the proper choice of constants. This figure makes it abundantly clear  that the equation of state parameter changes to negative values inside the proper range $(-1\leq\omega_{\Lambda}\leq0)$, which is in good agreement with astronomical data. As a result our research model is realistic. Fig. 6 shows that the equation of state parameter begins near to zero at the beginning of cosmic time (i.e., the universe is dominated by matter) and progresses to a close negative value of $-1$ at the end of cosmic time (i.e. when the Universe dominated by the HDE). Additionally, we can see that in the current universe, $\omega_ {\Lambda}$ tends to $-1$, indicating the model $\Lambda$CDM, whereas in the early universe, $-1<\omega_{\Lambda}<0$ suggests the quintessential model. Our model produces a $\omega_{\Lambda}=-0.90$ at the current epoch, which is near to the $\Lambda$CDM model i.e. $\omega_{\Lambda}=-1$  which is compatible with the observational bounds \cite{Planck18}. Fig. 7 displays the graphical behaviour of the Granda-Oliveros cutoff equation of state parameter of Renyi holographic dark energy vs time $t$ for the proper choice of constants. We observed that the value of $\omega_{\Lambda}$ is differed as compared to with the results obtained in the RHDE with Hubble cutoff. In this case, $\omega_{\Lambda}$ deviates from its initial positive value to function as a pure cosmological constant in the last phases of cosmic time. This model produces a $\omega_{\Lambda}=-0.72$ value at the current epoch, which is relatively close to the value produced by the $\Lambda$CDM model ($\omega_{\Lambda}=-1$), which is compatible with the observational bounds \cite{Planck18}.

\begin{figure}
\begin{center}
\includegraphics[scale=0.8]{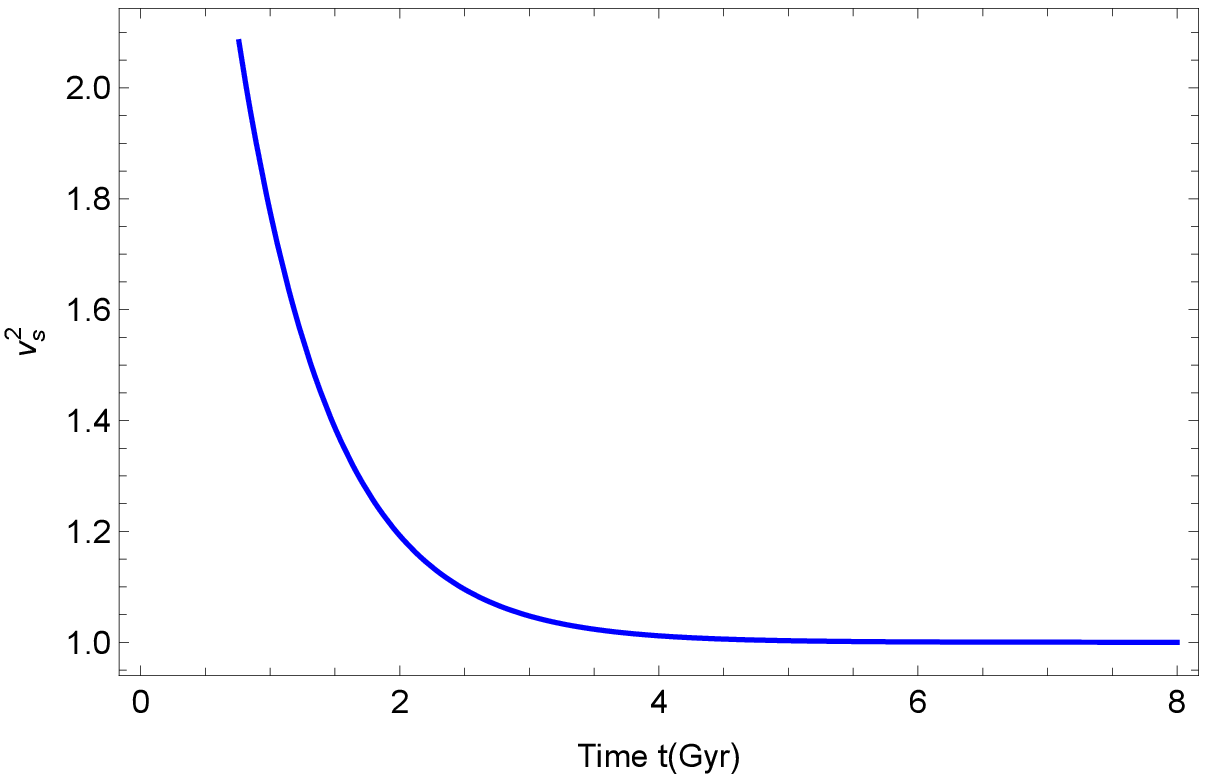}

\hspace{1cm}\vspace{5mm}
\footnotesize{Fig. 8. Square speed sound parameter ($v_{s}^{2}$) versus time ($t$)(Hubble horizon cutoff) for $\alpha=\beta=1.4$, $d=7$ and $\delta=5.2$.}
\vspace{3mm}
\end{center}
\end{figure}

\begin{figure}
\begin{center}
\includegraphics[scale=0.8]{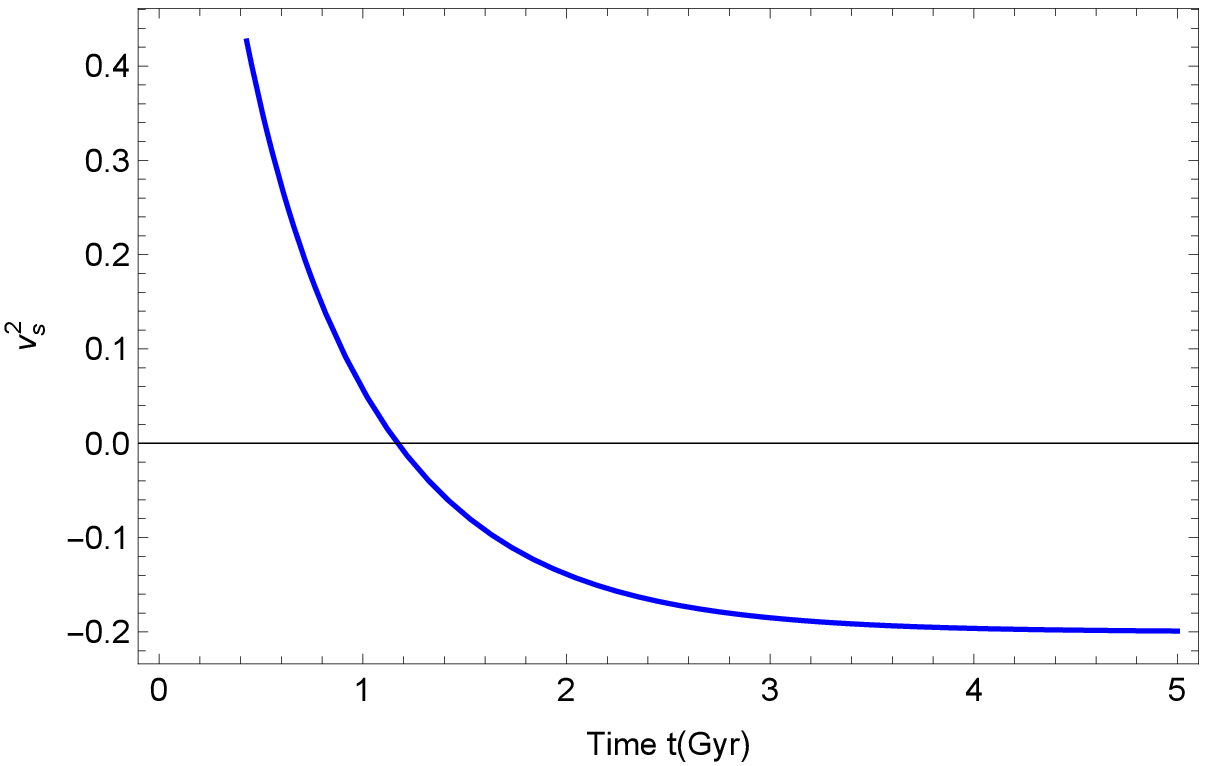}

\hspace{1cm}\vspace{5mm}
\footnotesize{Fig. 9. Square speed sound parameter ($v_{s}^{2}$) versus time ($t$)(GO cutoff) for $k_{1}=0.5, c_{1}=1, \alpha=\beta=1.4$, $d=7$, $\delta=5.2$, $\gamma_{1}=1.065$ and $\gamma_{2}=0.4$.}
\vspace{3mm}
\end{center}
\end{figure}

\subsection{Squared sound speed}
The squared sound speed parameter is given by
\begin{equation}\label{37}
v_{s}^{2}=\frac{\dot{p_{\Lambda}}}{\dot{\rho_{\Lambda}}}=\omega_{\Lambda}+\frac{\rho_{\Lambda}}{\dot{\rho_{\Lambda}}}\dot{\omega_{\Lambda}}.
\end{equation}
This parameter can be used to discuss how the stability of DE models is affected by its sign. If $v_{s}^{2}$ has a positive signature, the DE model is stable; otherwise, the model is unstable. Using equations \eqref{26},\eqref{35} and \eqref{34},\eqref{36} in the expression of squared sound speed $v_{s}^{2}$ equation \eqref{37}, we analyze $v_{s}^{2}$ graphically for both models -1 and 2. Fig. 8 displays the stability of RHDE with the Hubble cutoff for the proper choice of constants. It can be seen from the figure that the value of the $\delta$ has no effect on the stability of the universe. Also $v_{s}^{2}>0$ for all epoch and tends to a small value. Hence in all Universe our model is stable. Fig. 9 shows the stability of RHDE in the Granda-Oliveros cutoff of the model over time for the proper choice of constants. The model is stable during the beginning epoch, as can be seen in the figure. But after $t>1.17~Gyr$, the trajectory of the graph becomes negative, which shows that our model is classically unstable at current epoch.

 \subsection{Density parameter}
Total energy density parameter is given by
\begin{equation}
\Omega=\Omega_{m}+\Omega_{\Lambda},
\end{equation}
where $\Omega_{m}=\frac{\rho_{m}}{3H^{2}}$ is the matter density parameter and $\Omega_{\Lambda}=\frac{\rho_{\Lambda}}{3H^{2}}$ is the holographic dark energy density parameter. The total energy density parameters $\Omega>1$, $\Omega=1$, and $\Omega<1$ correspondingly represent the open, flat, and closed universes.
Now the total energy density parameter for RHDE with Hubble cutoff is found to be
\begin{eqnarray}
  \Omega &=& \frac{c_{1}k^{-3}(1+\alpha)^{2}(e^{\beta t}-1)^{\frac{-1+2\alpha}{1+\alpha}}}{3\beta^{2}e^{2\beta t}}+\frac{d^{2}\beta^{2}e^{2\beta t}}{\beta^{2}e^{2\beta t}+\pi\delta(1+\alpha)^{2}(e^{\beta t}-1)^{2}}.
\end{eqnarray}

And the total energy density parameter for RHDE with Granda-Oliveros cutoff is found as
\begin{equation}
\Omega=\frac{c_{1}k^{-3}(1+\alpha)^{2}(e^{\beta t}-1)^{\frac{-1+2\alpha}{1+\alpha}}}{3\beta^{2}e^{2\beta t}}+\frac{d^{2}\beta^{2}\{\gamma_{1}e{\beta t}-\gamma_{2}(1+\alpha)\}^{2}}{\pi \delta(1+\alpha)^{2}(e^{\beta t}-1)^{2}+\beta^{2}e^{\beta t}\{\gamma_{1}e{\beta t}-\gamma_{2}(1+\alpha)\}}.
\end{equation}

The total energy density parameter for the RHDE with the Hubble cutoff is shown in Fig. 10. Here, it is demonstrated that the energy density parameter's value was high in the early period of the universe but is currently approaching 1. So for very large times, the model predicts a flat universe. The resultant model is consistent with the observations because the universe as it currently exists is very close to flat. The total energy density parameter for RHDE with Granda-Oliveros cutoff is shown in Fig. 11. Here the graph is almost same as that of the Hubble cutoff. Hence the model predicts a flat Universe for large time.
\begin{figure}
\begin{center}
\includegraphics[scale=0.8]{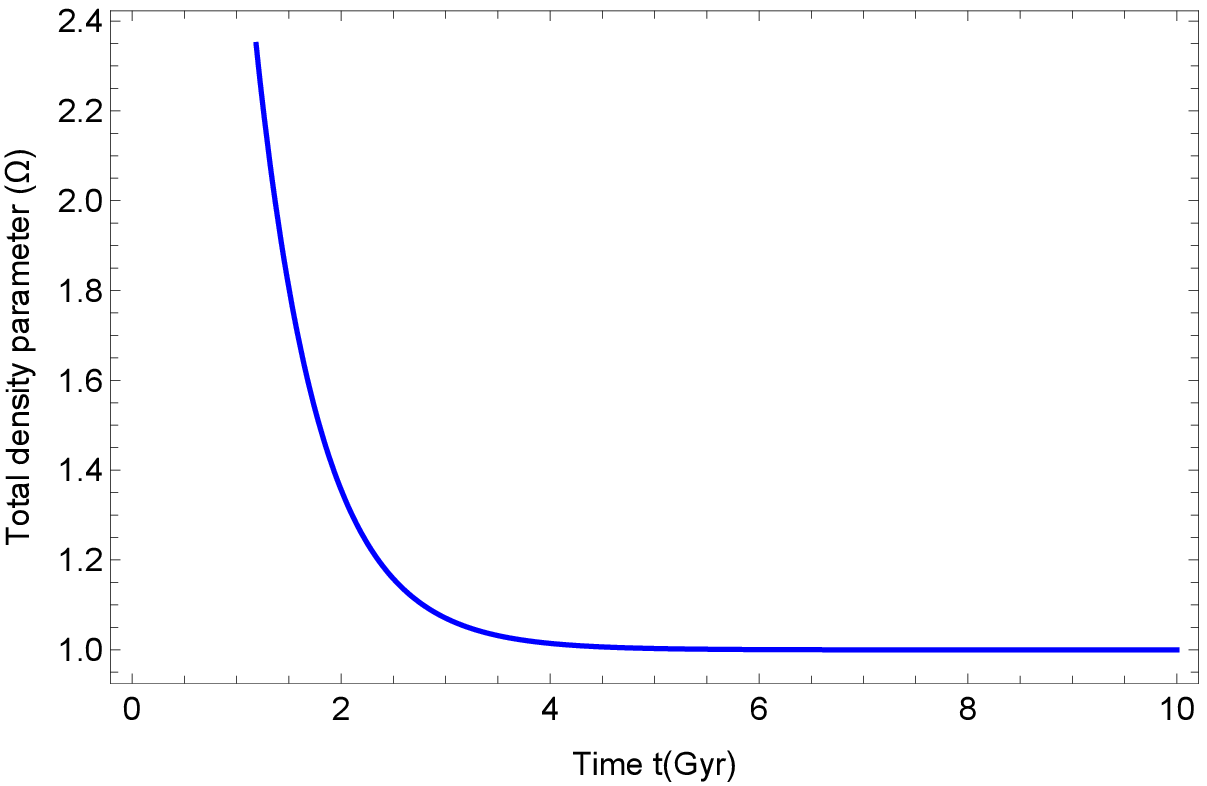}

\hspace{1cm}\vspace{5mm}
\footnotesize{Fig. 10.  Total energy density parameter($\Omega$) versus time ($t$)(Hubble horizon cutoff) for $k_{1}=0.5, c=1,\alpha=\beta=1.4$, $d=7$ and $\delta=5.2$.}
\vspace{3mm}
\end{center}
\end{figure}

\begin{figure}
\begin{center}
\includegraphics[scale=0.8]{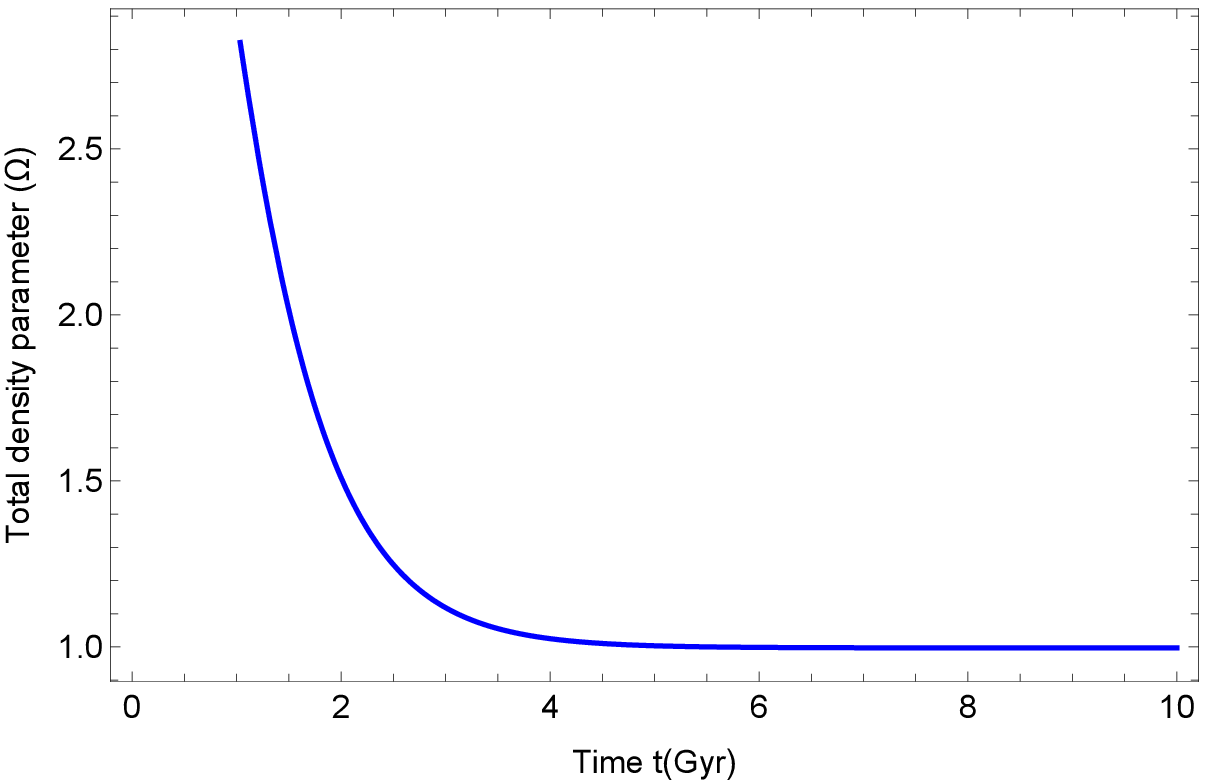}

\hspace{1cm}\vspace{5mm}
\footnotesize{Fig. 11. Total energy density parameter($\Omega$) versus time ($t$)(GO cutoff) for $k_{1}=0.5, c=1, \alpha=\beta=1.4$, $d=7$, $\delta=5.2$, $\gamma_{1}=1.065$ and $\gamma_{2}=0.4$.}
\vspace{3mm}
\end{center}
\end{figure}

\begin{figure}
\begin{center}
\includegraphics[scale=1]{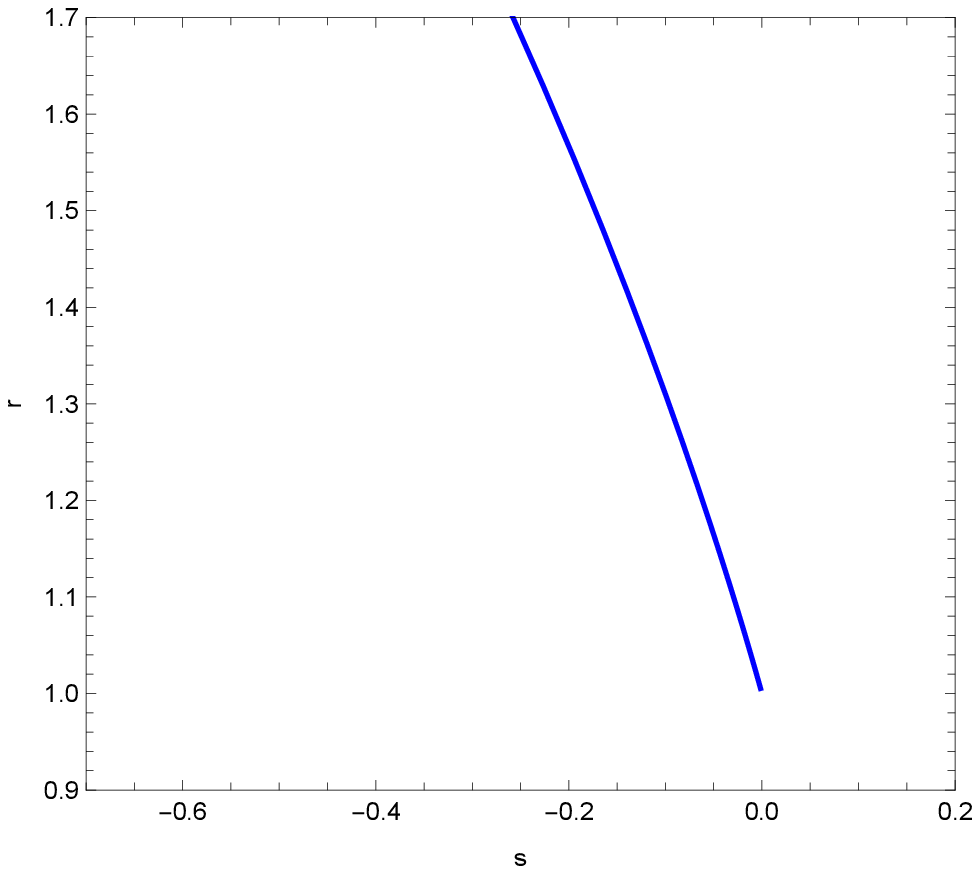}

\hspace{1cm}\vspace{5mm}
\footnotesize{Fig. 12. Plot of $r-s$ plane for $\alpha=\beta=1.4$, $d=7$.}
\vspace{3mm}
\end{center}
\end{figure}

\begin{figure}
\begin{center}
\includegraphics[scale=0.8]{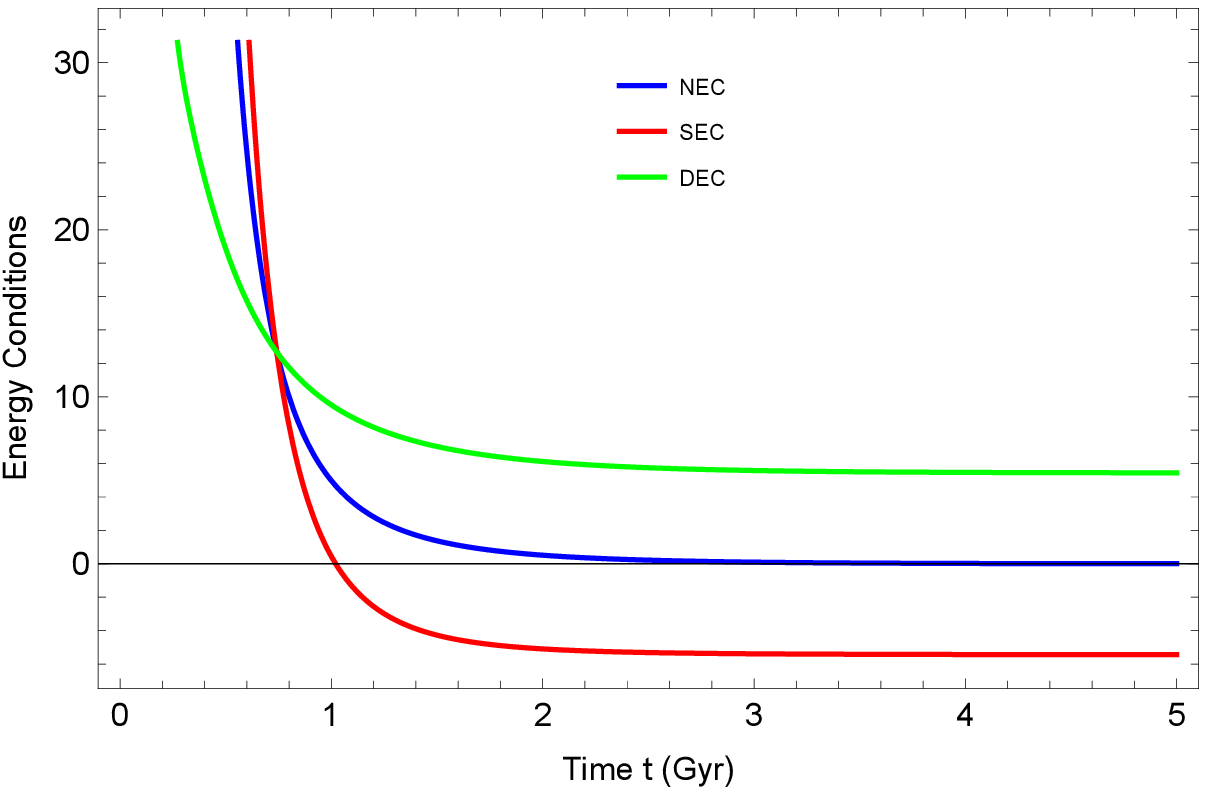}

\hspace{1cm}\vspace{5mm}
\footnotesize{Fig. 13. Energy conditions versus time ($t$)(Hubble horizon cutoff) for $\alpha=\beta=1.4$, $d=7$ and $\delta=5.2$.}
\vspace{3mm}
\end{center}
\end{figure}

\begin{figure}
\begin{center}
\includegraphics[scale=0.8]{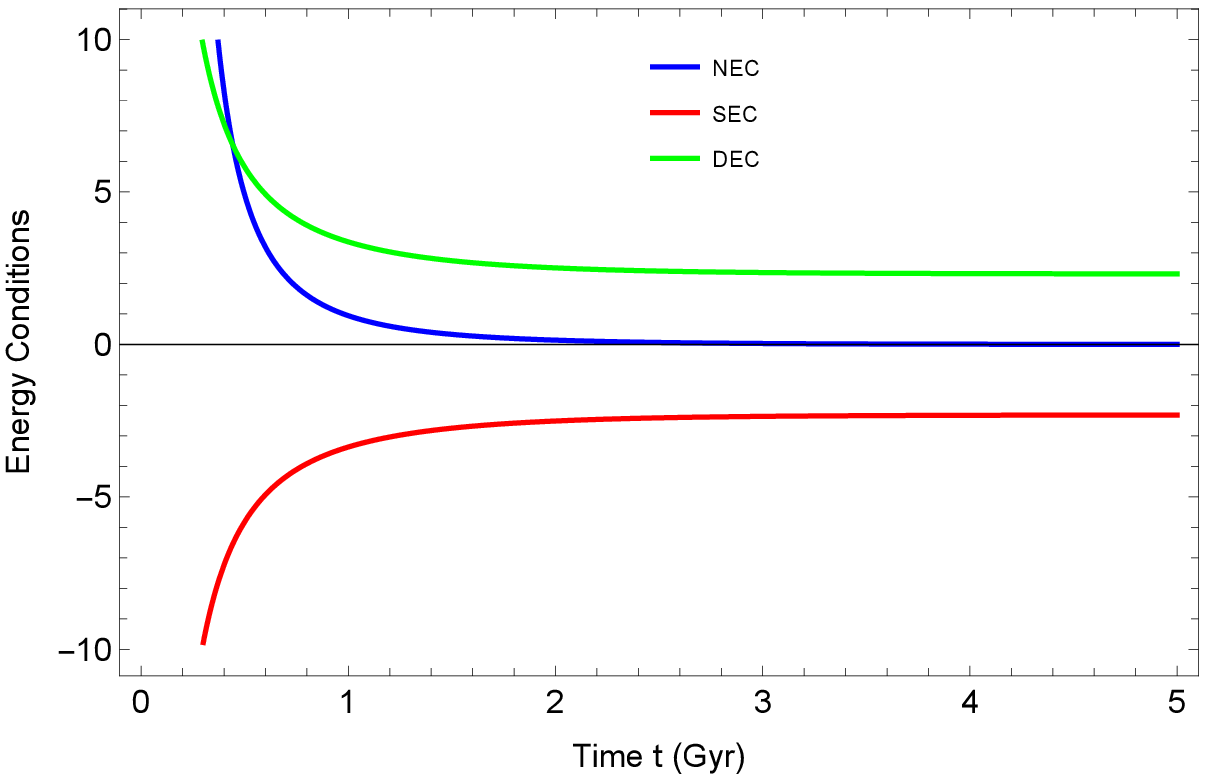}

\hspace{1cm}\vspace{5mm}
\footnotesize{Fig. 14. Energy conditions versus time ($t$)(GO horizon cutoff) for $\alpha=\beta=1.4$, $d=7$, $\delta=5.2$, $\gamma_{1}=1.065$ and $\gamma_{2}=0.4$.}
\vspace{3mm}
\end{center}
\end{figure}
\subsection{Statefinder parameters}
Hubble and deceleration parameters can be used to accurately explain the known universe expanding nature. The values of these parameters, however, are the same in many dynamical DE models at the present. As a result, these parameters were unable to choose the best-fitting model out of a variety of dynamical DE models. With this objective, Sahni et al. \cite{Sahni03} developed statefinder parameteres, which are dimensionless cosmological parameters and are defined as follows:
\begin{equation}
r=\frac{\dddot{a}}{aH^{3}},~~~~~~~~~s=\frac{r-1}{3(q-\frac{1}{2})}.
\end{equation}
For $(r,s)=(1,0)$ and $(r,s)=(1,1)$, respectively, these statefinders establish a connection with the $\Lambda$CDM and CDM models. In contrast to the Chaplygin gas model, which occurs for $r>1$ with $s<0$, if the trajectories of $r-s$ correspond to the region $s>0$ and $r<1$, the model belongs to the phantom and quintessence phases. These statefinders are same for both the models and are obtained as
\begin{equation}
r=1-\frac{3(1+\alpha)}{e^{\beta t}}+\frac{(1+\alpha)^{2}}{e^{2\beta t}}(e^{\beta t}+1),
\end{equation}
\begin{equation}
s=\frac{2(1+\alpha)[-3e^{\beta t}+(1+\alpha)(e^{\beta t}+1)]}{3e^{\beta t}[(1+\alpha)-3e^{\beta t}]}.
\end{equation}
Fig. 12 shows the graph of $(r,s)$ parameter in $r-s$ plane. The parameter $s$ is seen to remain negative for all values of $r$ at the early epoch. This suggests that the RHDE models were able to correspond to the Chaplygin gas model. Additionally, at late times, the $r-s$ plane corresponds to the $\Lambda$CDM.

\subsection{Energy conditions}
The energy conditions namely, null energy conditions (NEC), strong energy conditions (SEC) and dominant energy conditions (DEC), are respectively given by
(i) $\rho_{\Lambda}+p_{\Lambda}\geq0$,\\*
(ii) $\rho_{\Lambda}+3p_{\Lambda}\geq0$.\\*
(iii) $\rho_{\Lambda}-p_{\Lambda}\geq0$.\\*
Now the energy conditions for RHDE with Hubble cutoff are\\*
NEC:
\begin{eqnarray}\nonumber
\frac{3d^{2}(\beta e^{\beta t})^{4}}{\{\beta(1+\alpha)e^{\beta t}(e^{\beta t}-1)\}^{2}+\pi \delta\{(1+\alpha)(e^{\beta t}-1)\}^{4}}\\
\times \frac{2}{3}\frac{(1+\alpha)[\beta^{2}e^{2\beta t}+2\pi\delta (1+\alpha)^{2}(e^{\beta t}-1)^{2}]}{e^{\beta t}[\beta^{2}e^{2\beta t}+\pi\delta (1+\alpha)^{2}(e^{\beta t}-1)^{2}]}\geq0.
\end{eqnarray}
SEC:
\begin{eqnarray}\nonumber
\frac{3d^{2}(\beta e^{\beta t})^{4}}{\{\beta(1+\alpha)e^{\beta t}(e^{\beta t}-1)\}^{2}+\pi \delta\{(1+\alpha)(e^{\beta t}-1)\}^{4}}\\\times\bigg[\frac{2(1+\alpha)[\beta^{2}e^{2\beta t}+2\pi\delta (1+\alpha)^{2}(e^{\beta t}-1)^{2}]}{e^{\beta t}[\beta^{2}e^{2\beta t}+\pi\delta (1+\alpha)^{2}(e^{\beta t}-1)^{2}]}-2\bigg]\geq0.
\end{eqnarray}
DEC:
\begin{eqnarray}\nonumber
\frac{3d^{2}(\beta e^{\beta t})^{4}}{\{\beta(1+\alpha)e^{\beta t}(e^{\beta t}-1)\}^{2}+\pi \delta\{(1+\alpha)(e^{\beta t}-1)\}^{4}}\\\times\bigg[2-\frac{2(1+\alpha)[\beta^{2}e^{2\beta t}+2\pi\delta (1+\alpha)^{2}(e^{\beta t}-1)^{2}]}{3e^{\beta t}[\beta^{2}e^{2\beta t}+\pi\delta (1+\alpha)^{2}(e^{\beta t}-1)^{2}]}\bigg]\geq0.
\end{eqnarray}

Also the energy conditions for RHDE with Granda-Oliveros cutoff are found to be\\*
NEC:
\begin{eqnarray}\nonumber
\frac{3d^{2}\beta^{4} e^{2\beta t}\{\gamma_{1}e^{\beta t}-\gamma_{2}(1+\alpha)\}}{\pi\delta\{(1+\alpha)(e^{\beta t}-1)\}^{4}+e^{\beta t}\{\beta(1+\alpha)(e^{\beta t}-1)\}^{2}\{\gamma_{1}e^{\beta t}-\gamma_{2}(1+\alpha)\}}\\
\times \frac{(1+\alpha)\{2\gamma_{1} e^{\beta t}-\gamma_{2}(1+\alpha)(e^{\beta t}+1)\}}{3e^{\beta t}\{\gamma_{1}e^{\beta t}-\gamma_{2}(1+\alpha)\}} \frac{[2\pi\delta(1+\alpha)^{2}(e^{\beta t}-1)^{2}+\beta^{2}e^{\beta t}\{\gamma_{1}e^{\beta t}-\gamma_{2}(1+\alpha)\}]}{[\pi\delta\{(1+\alpha)(e^{\beta t}-1)\}^{2}+\beta^{2}e^{\beta t}\{\gamma_{1}e^{\beta t}-\gamma_{2}(1+\alpha)\}]}\geq0.
\end{eqnarray}
SEC:
\begin{eqnarray}\nonumber
\frac{3d^{2}\beta^{4} e^{2\beta t}\{\gamma_{1}e^{\beta t}-\gamma_{2}(1+\alpha)\}^{2}}{\pi\delta\{(1+\alpha)(e^{\beta t}-1)\}^{4}+e^{\beta t}\{\beta(1+\alpha)(e^{\beta t}-1)\}^{2}\{\gamma_{1}e^{\beta t}-\gamma_{2}(1+\alpha)\}}\\
\times \bigg[\frac{(1+\alpha)\{2\gamma_{1} e^{\beta t}-\gamma_{2}(1+\alpha)(e^{\beta t}+1)\}}{e^{\beta t}\{\gamma_{1}e^{\beta t}-\gamma_{2}(1+\alpha)\}} \frac{[2\pi\delta(1+\alpha)^{2}(e^{\beta t}-1)^{2}+\beta^{2}e^{\beta t}\{\gamma_{1}e^{\beta t}-\gamma_{2}(1+\alpha)\}]}{[\pi\delta\{(1+\alpha)(e^{\beta t}-1)\}^{2}+\beta^{2}e^{\beta t}\{\gamma_{1}e^{\beta t}-\gamma_{2}(1+\alpha)\}]}-2\bigg]\geq0.
\end{eqnarray}
DEC:
\begin{eqnarray}\nonumber
\frac{3d^{2}\beta^{4} e^{2\beta t}\{\gamma_{1}e^{\beta t}-\gamma_{2}(1+\alpha)\}^{2}}{\pi\delta\{(1+\alpha)(e^{\beta t}-1)\}^{4}+e^{\beta t}\{\beta(1+\alpha)(e^{\beta t}-1)\}^{2}\{\gamma_{1}e^{\beta t}-\gamma_{2}(1+\alpha)\}}\\
\times \bigg[2-\frac{(1+\alpha)\{2\gamma_{1} e^{\beta t}-\gamma_{2}(1+\alpha)(e^{\beta t}+1)\}}{3e^{\beta t}\{\gamma_{1}e^{\beta t}-\gamma_{2}(1+\alpha)\}} \frac{[2\pi\delta(1+\alpha)^{2}(e^{\beta t}-1)^{2}+\beta^{2}e^{\beta t}\{\gamma_{1}e^{\beta t}-\gamma_{2}(1+\alpha)\}]}{[\pi\delta\{(1+\alpha)(e^{\beta t}-1)\}^{2}+\beta^{2}e^{\beta t}\{\gamma_{1}e^{\beta t}-\gamma_{2}(1+\alpha)\}]}\bigg]\geq0.
\end{eqnarray}
Fig. 13 shows the graph of energy conditions for RHDE with Hubble cutoff for our model. From the graph, it is observed that $\rho_{\Lambda}+p_{\Lambda}\geq0$ and $\rho_{\Lambda}-p_{\Lambda}>0$ but $\rho_{\Lambda}+3p_{\Lambda}\geq0$ at early times but becomes negative after some time and stays in the negative domain. So, NEC and DEC are satisfied whereas SEC is violated. Fig. 14 shows the graph of energy conditions for RHDE with Granda-Oliveros cutoff for our model. From the graph, it is observed that $\rho_{\Lambda}+p_{\Lambda}\geq0$ and $\rho_{\Lambda}-p_{\Lambda}>0$ but $\rho_{\Lambda}+3p_{\Lambda}<0$. This shows that NEC and DEC are satisfied whereas SEC is violated. So in both the model NEC and DEC are satisfied whereas SEC is violated in the present
and future. Therefore, the universe accelerates as a result of the SEC violation. Our model shows the shift from an early decelerating to a current accelerating universe as a result of the violation of SEC, which causes an anti-gravitational effect that causes the universe to jerk. Our model therefore fits the most recent cosmological observations.

\section{Conclusion}\label{sec6}
In this work, we investigate RHDE with a homogeneous and anisotropic Universe of Bianchi type-I, in the context of $f(G)$ gravity. We also consider  RHDE with the IR cutoffs of both the Hubble and the Granda-Oliveros horizons. We make the assumption that the deceleration parameter (DP) is a function of Hubble parameter $H$ in order to determine exact solutions to the field equations. With the use of this analysis, we found that the deceleration parameter changes from negative to positive with respect to redshift $z$, indicating that the universe transitions from an earlier deceleration phase to the present acceleration phase. Our model's transition redshift value is $z_{tr}=0.73$, which is in accordance with the observational data. Scalar expansion and shear scalar both have infinitely large value at $t\rightarrow0$ and become finite at $t\rightarrow\infty$. Since the anisotropic parameter doesn't change throughout the cosmic evolution, our model is fully anisotropic from the early Universe to the end of the Universe for $m\neq1$ whereas the model is isotropic for $m=1$. For investigations in model I, it has been found that the energy density of the model is consistently a positive function of time, and that these parameters have no effect on the behavior of the model for any $\alpha\geq0.3$. Also, the RHDE density in Hubble's cutoff is positive for all Universe and is decreasing to a small value at at later times.  The RHDE universe in the Hubble's cutoff is stable, and the value of $\delta$  has no effect on the stability of the universe, which is approaching to a small value. From the evolution of the EoS parameter, we understand that in the early universe, it indicates the quintessential model, while in the current universe, $\omega_{\Lambda}$ tends to $-1$, i.e. the model $\Lambda$CDM, which is well in agreement with recent observational data. Additionally, the NEC and DEC energy conditions are satisfied, however the SEC is violated at later times. The acceleration of the universe results from this SEC violation. Again in the study of model II, the energy density of the model is rigorously a positive function of time and is a decreasing function and approaches to a small positive value at later times. Even if it is stable in the early Universe, the behavior of the stability of the RHDE Universe in the Granda-Oliveros cutoff is not stable at later times. In this model, the EoS parameter falls from a positive value in the early phase of cosmic time to act as a pure cosmological constant, or $\omega_{\Lambda}=-1$, in the late phase. The NEC and DEC energy conditions are also satisfied, while the SEC is violated in the present and the future, which causes the Universe to accelerate. Additionally, for both models, the $(r,s)$ plane provides a correspondence with the Chaplygin gas model and, at late times, with the $\Lambda$CDM. Finally, the exact solutions described in the study can be one of the decent candidates to describe the observable Universe. In order to comprehend the characteristics of the anisotropic Bianchi type-I model in the development of the Universe, it may be helpful to consider the solutions presented in this study.

\end{document}